\documentclass[12pt]{article}
\usepackage{amsfonts}
\usepackage[dvips]{epsfig}
\usepackage{graphicx}
\begin{document}

\title{Dynamical Systems, Topology and Conductivity in
Normal Metals in strong magnetic fields.}

\author{A.Ya.Maltsev$^{1}$, S.P.Novikov$^{1,2}$.}

\date{
\centerline{$^{(1)}$ L.D.Landau Institute for Theoretical Physics,}
\centerline{119334 ul. Kosygina 2, Moscow, }
\centerline{ maltsev@itp.ac.ru \,\, ,
\,\,\, novikov@itp.ac.ru}
\centerline{$^{(2)}$ IPST, University of Maryland,}
\centerline{College Park MD 20742-2431,USA}
\centerline{novikov@ipst.umd.edu}}

\maketitle

\begin{abstract}
 We represent here the full description of all asymptotic regimes
of conductivity behavior in the so-called "Geometric Strong
Magnetic Field limit" in the  3D single crystal normal metals with
topologically complicated Fermi surfaces. In particular, new
observable integer-valued characteristics of conductivity of the
topological origin were introduced by the present authors few
years ago; they are based on the Topological Resonance found by
the present authors and play the basic role in the total picture.
Our investigation is based on the study of dynamical systems on
Fermi surfaces for the semi-classical motion of electron in
magnetic field realized by the Moscow topological group.
\end{abstract}

\vspace{0.5cm}

{\it This paper is the reduced and improved version of the paper
cond-mat/0304471. The authors are very grateful to 
Prof. Joel L. Lebowitz for his help and interest to this work.}

\vspace{0.5cm}

\section{Introduction.}

 We are going to consider the  implications of the so called
 "Geometric
Strong Magnetic Field limit" for the conductivity in normal metals
with topologically complicated Fermi surface in the presence of
the homogeneous magnetic field. The corresponding limit can be
defined by the relation $\omega_{B} \tau \gg 1$. Here
$\omega_{B}$ is the cyclotron frequency for the electron in
crystal and $\tau$ is the mean free motion time between the
scattering acts. This means actually that this theory is based on
the Kinetic equation for the electron gas in crystal for the
semiclassical  electrons  in the external fields. Let us say that
the corresponding conditions for the external fields are always
satisfied for the experimentally available electric and magnetic
fields in the case of normal metals. We can speak, for example,
about the limit of very strong magnetic fields in the experimental
sense where the semiclassical approximation still gives the main
features of transport phenomena. It works until the magnetic flux
through the elementary cell of the ion lattice is small in
comparison with the quantum unit. Taking into account the value of
the physical parameters in the real single crystal normal metal
(like gold, for example) we have finally $1t \ll B\ll 10^3t$ for
the temperatures like $T \leq 1K$. We will not discuss here
any questions of rigorous foundations of this approach (very
standard in the physics literature dedicated to the transport
phenomena). The detailed explanations of this method can be found
in classical books (see, for example 
\cite{kittel,etm,ziman,abrikosov}). 
Let us give here also the references 
\cite{PanSpTeuf1,PanSpTeuf2} where
the mathematically rigorous approach to the semiclassical motion
of electron in electromagnetic field and lattice as well as the
historical remarks can be found. Indeed, no rigorous theory
of the Kinetic equation here was developed yet, so these papers
don't make our results about conductivity more rigorous.

 We will consider the electron states in crystal parameterized by the
energy bands and the quasimomentum ${\bf p}$ defined modulo the
reciprocal lattice vectors. From the topological point of view we
can say that  quasimomentum belongs to the three-dimensional torus
${\mathbb T}^{3}$ (Brillouen zone) rather than to the Euclidean
space ${\mathbb R}^{3}$. The torus ${\mathbb T}^{3}$ arises as
factorization of the space ${\mathbb R}^{3}$ with respect to the
reciprocal lattice. Topologists say that the space ${\mathbb
R}^{3}$ is a covering over 3-torus ${\mathbb T}^{3}$. The periodic
dispersion relation $\epsilon({\bf p})$ of any energy band can be
considered as the one-valued continuous function on the torus
${\mathbb T}^{3}$. The  Fermi surface $S_{F}: \epsilon({\bf p}) =
\epsilon_{F}$ can also be considered as the smooth compact
two-dimensional surface without boundary embedded in the
three-dimensional torus ${\mathbb T}^{3}$. In this paper we will
often compare these two pictures in the 3-torus ${\mathbb T}^{3}$
and in the total Euclidean 3-space of quasimomenta. We will use
the equation ${\dot {\bf p}} = {\bf F}_{ext}$ both in the torus
${\mathbb T}^{3}$ and in the covering Euclidean 3-space ${\mathbb
R}^{3}$ for the homogeneous force ${\bf F}_{ext}$. In particular,
we will consider the properties of the trajectories of this system
both in these two spaces which will be very convenient for our
consideration. We apologize for using topological terminology in
some parts of this paper. However we could not avoid it.

  Following the standard approach, we consider a system:

$${\dot {\bf p}} = {e \over c} \left[\nabla \epsilon({\bf p})
\times {\bf B} \right] + e {\bf E} $$ for the semiclassical
electron in both homogeneous electric and magnetic field. The
value of electric field $E$ is going to be infinitesimally small
in measuring the conductivity.  Therefore only the trajectories of
the main dynamical system

\begin{equation}
\label{dynsyst}
{\dot {\bf p}} = {e \over c}
\left[\nabla \epsilon({\bf p}) \times {\bf B} \right]
\end{equation}
should be investigated in this approach.

 The trajectories
of (\ref{dynsyst}) in the Euclidean 3-space are given on every
energy level $\epsilon({\bf p}) = const$ by the  plane sections
orthogonal to magnetic field. So we have  the analytic
integrability of the system (\ref{dynsyst}) in the 3-space $R^3$.
However, the global structure of the trajectories on the 3-torus
can be highly nontrivial after identification  the quasimomenta
equivalent modulo the reciprocal lattice.

 The dynamical system (\ref{dynsyst}) conserves also the volume
element $d^{3} p$ in
${\mathbb T}^{3}$ and does not change at all the
Fermi distribution. So, in the absence of the
electric field ${\bf E}$ we will have the electron distribution
unchanged (up to the quantum corrections). Nevertheless, the
response of this system to small perturbations will be completely
different from the case ${\bf B} = 0$ and depend strongly on the
geometry of trajectories of the dynamical system (\ref{dynsyst}).

 Let us say that this dependence was first discovered by the
school of I.M. Lifshitz (I.M. Lifshitz, M.Ya. Azbel, M.I. Kaganov,
V.G. Peschanskii
\cite{lifazkag,lifpes1,lifpes2,lifkag1,lifkag2,etm}) in 1950's.
Thus, in the work \cite{lifazkag} the crucial difference in
conductivity was found for the contribution of the closed and open
periodic electron trajectories in ${\bf p}$-space considered as
the total Euclidean 3-space ${\mathbb R}^{3}$. Namely, it was
shown that the first case corresponds to the total decreasing of
conductivity in the plane orthogonal to ${\bf B}$ for $B
\rightarrow \infty$ while the second case corresponds to the
strong anisotropy of conductivity in the plane orthogonal to ${\bf
B}$ in the same limit:  the conductivity vanishes just in one
direction in this plane only depending on the mean direction of
the open periodic trajectory. In the works \cite{lifpes1,lifpes2}
the interesting examples of Fermi surfaces and  electron
trajectories were considered. However the work \cite{lifpes2}
contains some conceptual mistake: open trajectories were found for
the generic family of magnetic fields with different mean
directions. This result is wrong. It contradicts to the
"Topological Resonance" which is a base of our main results
\cite{novmal1,novmal2}. We will discuss it in the Chapter 2.

 The problem of classification of all possible trajectories
on the Fermi surfaces was first set by S.P. Novikov
(\cite{novikov1}) and then considered in his school (S.P.
Novikov,A.V. Zorich, I.A. Dynnikov, S.P. Tsarev, A.Ya. Maltsev
\cite{zorich1,novikov2,dynn1,dynn2,dynn3,tsarev,novikov3,zorich2,
novikov4,novmal1,dynn5,dynn4,dynmal,maltsev,novmal2,dynn7,novikov5,
DeLeo,malnov3,malnov4,maltsev2}).

The full classification of all possible situations in the
Geometric Strong Magnetic Field Limit (GSMF-limit) can be given as
a result of the topological studies of this important class of
dynamical systems on 2-dimensional surfaces. The most important
feature of this new picture is the invention of the observable
"Topological numbers" in the conductivity which always appear in
GSMF-limit in the situation when the conductivity in the plane
orthogonal to the generic magnetic field ${\bf B}$ reveals the
strong anisotropy for $B \rightarrow \infty$ which is stable with
respect to the small rotations of the directions of ${\bf B}$.
These Topological numbers have the form of the triples of integers
$(m_{1}^{\alpha}, m_{2}^{\alpha}, m_{3}^{\alpha})$. They  are
connected with some  integral planes $\Gamma^{\alpha}$ in the
reciprocal lattice. The corresponding directions of ${\bf B}$ for
which the given triple $(m_{1}^{\alpha}, m_{2}^{\alpha},
m_{3}^{\alpha})$ can be observed give always a region
$\Omega_{\alpha}$ of non-zero measure (on the unit sphere) among
the total set of directions of ${\bf B}$. We call the
corresponding region $\Omega_{\alpha}$ on the unit sphere the
"Stability zone" corresponding to given triple $(m_{1}^{\alpha},
m_{2}^{\alpha}, m_{3}^{\alpha})$ due to the topological "rigidity"
of the triple $(m_{1}^{\alpha}, m_{2}^{\alpha}, m_{3}^{\alpha})$
within $\Omega_{\alpha}$.

 We claim also that there are only two types of the stable conductivity
tensor asymptotic  behavior in the GSMF-limit ($B \rightarrow
\infty$) for any normal metal with arbitrary topologically
complicated Fermi surface. Namely, the (''topologically
nontrivial'') case of the strongly anisotropic behavior of
conductivity in the plane orthogonal to ${\bf B}$ corresponding to
some triple of Topological numbers and the (''trivial'') case of
the uniform decreasing of conductivity in any direction orthogonal
to ${\bf B}$ for $B \rightarrow \infty$. These cases cover the
area on the 2-sphere of the full total measure, so the generic
directions are either of the first type or of the second type.

 All other types of conductivity
behavior in GSMF-limit can not be stable with respect to small rotations
of ${\bf B}$. We don't give in this paper all topological proofs of these
facts because of their rather high mathematical complexity and give just the
corresponding citations on the mathematical and physical literature
(\cite{novikov1,zorich1,novikov2,dynn1,dynn2,dynn3,tsarev,novikov3,
zorich2,novikov4,novmal1,dynn5,dynn4,dynmal,maltsev,novmal2,dynn7,
novikov5,DeLeo,malnov3,malnov4,maltsev2}. Let us introduce here two
notations for these stable situations which we will use in this paper.

 {\bf Situation A.}(Topologically trivial behavior).

We call it the situation A; it is
 the case of uniform decreasing
of conductivity in the plane orthogonal to ${\bf B}$ for $B
\rightarrow \infty$.

 {\bf Situation B.} (Topological numbers and Topological 
Resonance).

This is the case of the strong anisotropy of conductivity in the
plane orthogonal to ${\bf B}$ with decreasing in just one
direction in this plane for $B \rightarrow \infty$. This direction
can be described as the intersection of the plane orthogonal to
magnetic field with some integral plane (given by two reciprocal
lattice vectors). The corresponding integral plane remains unchanged 
under the small rotations
of the magnetic field. Three integers characterizing this plane in
the reciprocal lattice are exactly the observable topological
numbers. Topological Resonance claiming that the mean directions
of all open trajectories coincide for the generic magnetic field
is a base of this result. It was extracted by the present authors
from the core of the topological works quoted above. As it was
already mentioned, the conceptual mistake has been made exactly
here in the classical works of the Lifshitz group.

 As we just said above all the stable cases in GSMF-limit can be
described just by two situations. However, for complicated enough
Fermi surfaces also the rather nontrivial "chaotic" behavior of
conductivity tensor in GSMF-limit is possible
(\cite{maltsev,novmal2, malnov3,malnov4}) for the set of
directions of the zero measure. The trajectories of this type were
completely unknown in classical literature. They were discovered
recently in the topological works (\cite{tsarev,dynn4,dynn7}).
Chaotic trajectories can be divided into two different classes:

 1)Weakly  chaotic trajectories (the Tsarev type);

 2)Strongly chaotic trajectories  (the Dynnikov type).

 Let us say here some words about these two classes.

 The trajectories of the
first kind can appear only if the direction of ${\bf B}$ is
"partly rational", i.e. the plane $\Pi({\bf B})$ orthogonal to
${\bf B}$ contains one (up to the multiplier) reciprocal lattice
vector. The trajectories of the second kind can appear only if the
direction of $({\bf B})$ is completely irrational, i.e. $\Pi({\bf
B})$ does not contain any reciprocal lattice vectors. In the case
when the direction of ${\bf B}$ is purely rational (i.e. $\Pi({\bf
B})$ contains two linearly independent reciprocal lattice vectors)
the chaotic electron trajectories can not appear.

 The corresponding behavior of conductivity in GSMF-limit is also very
different for these two classes of chaotic electron trajectories
(\cite{maltsev,novmal2,malnov3,malnov4}). Thus in the case of
weakly  chaotic trajectories  the asymptotic expression of
conductivity is just slightly different from the Situation B and
corresponds also to strongly anisotropic behavior of conductivity
in the plane $\Pi({\bf B})$ for $B \rightarrow \infty$
(\cite{malnov3,malnov4}). Nevertheless, this regime is unstable
with respect to the small rotations of ${\bf B}$ unlike the
regular case where the "Stability zones" can be observed.

 The strongly chaotic trajectories, however, demonstrates completely
different behavior of $\sigma^{ik}(B)$ in GSMF-limit (\cite{maltsev}).
Namely, in this case the conductivity in the plane orthogonal to ${\bf B}$
decreases as $B \rightarrow \infty$ (in all directions) with the different
from the Situation A analytic dependence on $B$. Besides that, in this case
the part of the Fermi surface is excluded also from the conductivity along the
direction of ${\bf B}$ for $B \rightarrow \infty$. The last fact leads to the
"sharp minimum" in the conductivity along ${\bf B}$ for the given direction
of ${\bf B}$ if the chaotic trajectories of the second kind appear. Usually
the conductivity along ${\bf B}$ remains finite in this "sharp minima" since
only a part of the Fermi surface becomes excluded from the corresponding
contribution. However, these minima can be observed (on the unit sphere)
using the small rotations of ${\bf B}$ due to the instability of all the
chaotic trajectories with respect to such rotations.

 In Chapter 3 we give more detailed description of the chaotic trajectories
with the corresponding consideration of conductivity.

\section{Topologically rigid cases and Topological numbers.}

 Let us define the "Degree of irrationality" of
magnetic field with respect to reciprocal lattice.

{\it Let $\{{\bf g}_{1}, {\bf g}_{2}, {\bf g}_{3}\}$ be the basis
of the reciprocal lattice $\Gamma^{*}$ such that the vectors of
$\Gamma^{*}$ are given by all possible integer linear
combinations of $\{{\bf g}_{1}, {\bf g}_{2}, {\bf g}_{3}\}$. Then:

 1) The direction of ${\bf B}$ is rational (or has irrationality $1$)
if the plane $\Pi({\bf B})$ orthogonal to ${\bf B}$ contains two
linearly independent reciprocal lattice vectors.

 2) The direction of ${\bf B}$ has irrationality $2$ if the plane
$\Pi({\bf B})$ contains just one (up to multiplier) reciprocal
lattice vector.

 3) The direction of ${\bf B}$ has irrationality $3$ (or completely
irrational) if the plane $\Pi({\bf B})$ does not contain any
reciprocal lattice vectors.
}

 The generic directions of magnetic field are completely
 irrational.
The direction of ${\bf B}$ should be "specially chosen" to have
irrationality 1 or 2. We are going to consider now the situations
stable with respect to the small rotations of ${\bf B}$. This
means in particular that  specific features of such cases should
not be connected with any kind of rationality of the direction of
${\bf B}$, i.e. they should reveal all their properties for the
completely irrational directions of magnetic field.

 The electron trajectories
are given by the intersections of the periodic Fermi surface with
the family of parallel planes orthogonal to the magnetic field.
For simplicity we will assume in this Chapter that the direction
of ${\bf B}$ is completely irrational (for example,  no open
periodic trajectories can appear in the planes orthogonal to ${\bf
B}$). Let us postpone the specific (unstable) features of rational
directions to the next Chapter.

{\it We call the trajectory non-singular if it
is not adjacent to the critical point. The trajectories adjacent to
the critical points as well as the critical points themselves we call
singular trajectories.}

{\it We call the non-singular trajectory compact
if it is closed on the plane. We call the non-singular trajectory
open if it is unbounded in ${\mathbb R}^{2}$.}

\begin{figure}
\begin{center}
\epsfig{file=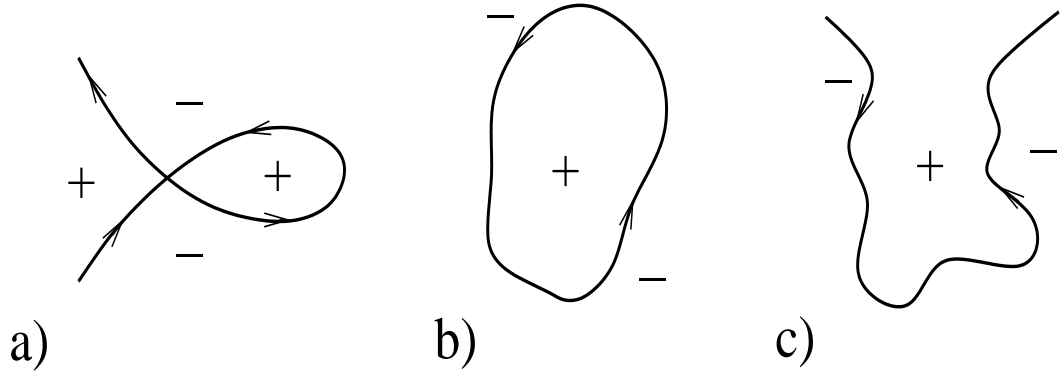,width=14.0cm,height=5cm}
\end{center}
\caption{The singular, compact and open non-singular
 trajectories. The signs $"+"$ and $"-"$ show the
regions of larger and smaller values of
$\epsilon({\bf p})|_{\Pi}$ respectively.
}
\label{traject}
\end{figure}

 The examples of singular, compact and open non-singular
trajectories are shown on the Fig. \ref{traject}, a-c.

 We give now the important definitions concerning the behavior
of open trajectories in  the planes orthogonal to ${\bf B}$. It is
 this type of trajectories which  plays the main role in the
GSMF-limit of conductivity.

\begin{figure}
\begin{center}
\epsfig{file=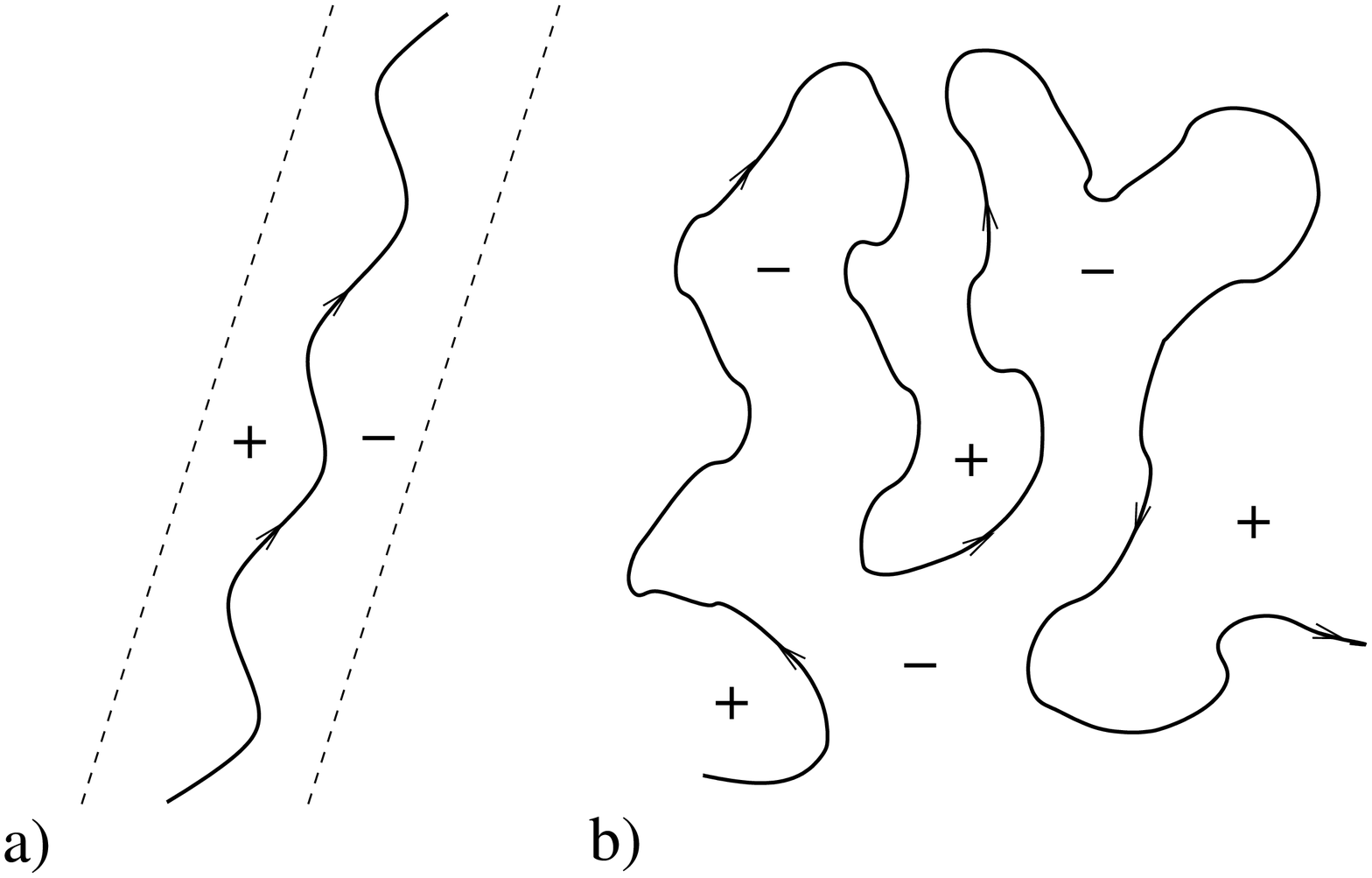,width=14.0cm,height=7cm}
\end{center}
\caption{"Topologically regular" (a) and "chaotic"
(b) open trajectories in the plane $\Pi$ orthogonal to ${\bf B}$.
}
\label{regandch}
\end{figure}

{\it We call the open trajectory topologically regular (i.e.
corresponding to the "topologically integrable" case) if it lies
within the straight line strip of the finite width in ${\mathbb
R}^{2}$ and passes through it from $-\infty$ to $\infty$ (see Fig.
\ref{regandch}, a). All other open trajectories we  call chaotic
(Fig. \ref{regandch}, b).}

 Let us point out that the topologically regular open trajectories
are not periodic at all which would contradict to the
irrationality of the direction of ${\bf B}$. In fact they are
quasiperiodic.\footnote{The ergodic properties of trajectories
of this kind were investigated in \cite{arnold,sinkhan}} The 
property of topological regularity is connected
with the rather nontrivial property of the "Carriers" of such
trajectories in the ${\bf p}$-space which we are going to consider
below. Let us just say now that this property is closely connected
with the "Topological numbers" which we are going to introduce.

\begin{figure}
\begin{center}
\epsfig{file=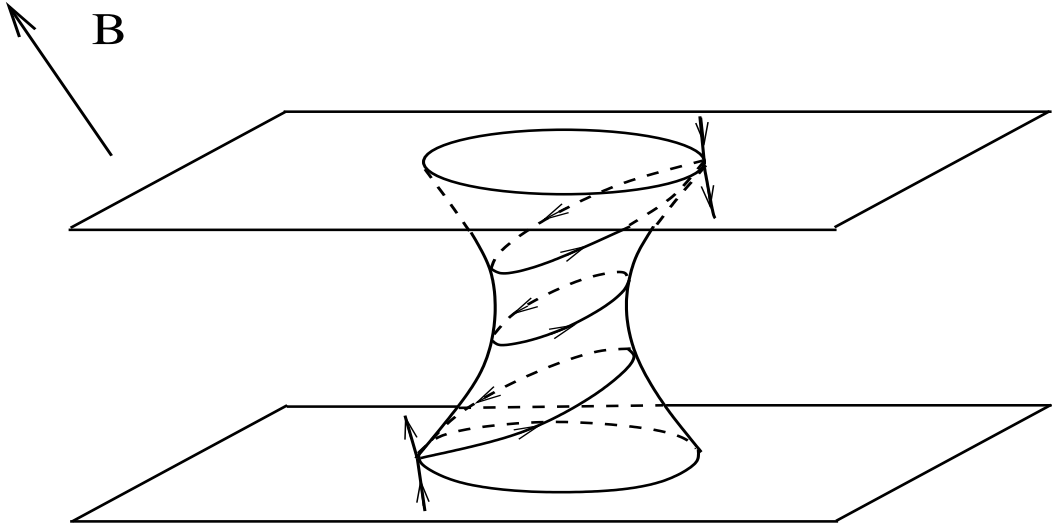,width=14.0cm,height=7cm}
\end{center}
\caption{The cylinder of compact trajectories bounded by the singular
orbits.(The simplest case of just one critical point on the singular
trajectory.)
}
\label{cylcltr}
\end{figure}

 Let us introduce now the "Carriers of open trajectories"
and the "Topological numbers".  We follow here the convenient
description (\cite{dynn4},\cite{dynn7})  of the Fermi surface with
the trajectories  on it when the direction of ${\bf B}$ is fixed.
We will be interested first of all in the open electron
trajectories in the ${\bf p}$-space.  Let us say that in general
just a part of the Fermi surface will be covered by the open
electron trajectories. The remaining part will contain compact (or
singular) trajectories. Let us remove  all parts of the Fermi
surface covered by the non-singular compact trajectories. The
remaining part

$$S_{F}/({\rm Compact \, Nonsingular \, Trajectories}) \,\,\, =
\,\,\, \cup_{j} \, S_{j}$$
is a union of the $2$-manifolds $S_{j}$ with boundaries
$\partial S_{j}$ who are the compact singular trajectories.
The generic type in this case is a separatrix orbit with just one
critical point like on the Fig. \ref{cylcltr}.

\vspace{0.5cm}

{\it We call every piece $S_{j}$ the
{\bf "Carrier of open trajectories"}.}

\vspace{0.5cm}

\begin{figure}
\begin{center}
\epsfig{file=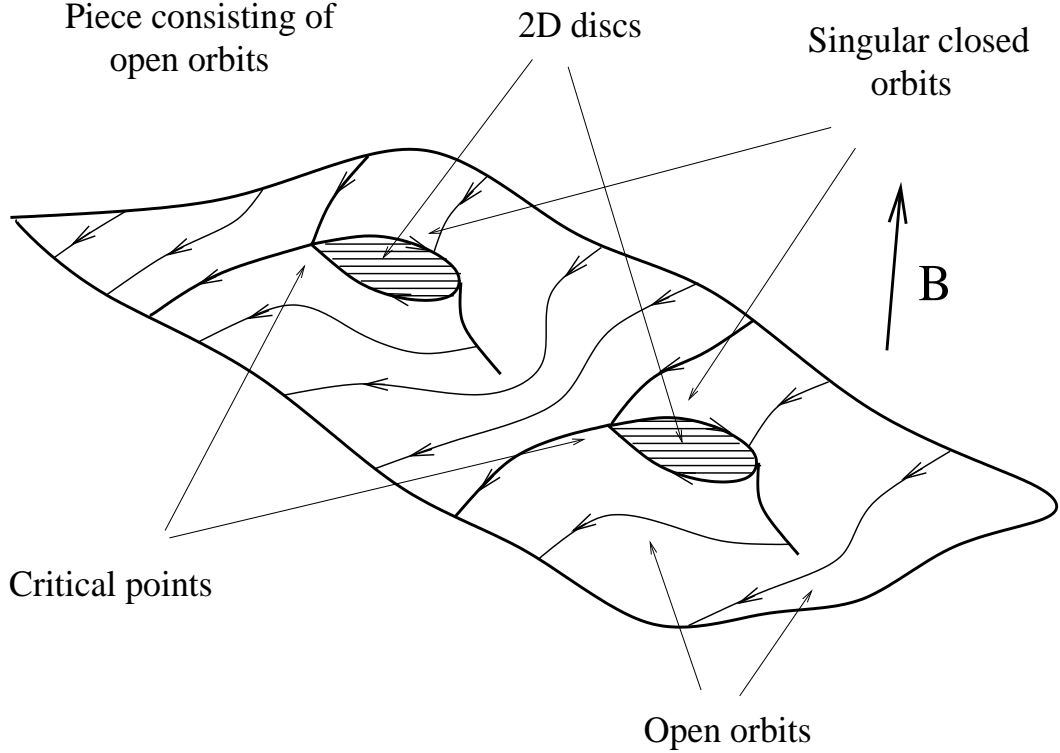,width=14.0cm,height=10cm}
\end{center}
\caption{The reconstructed constant energy surface with removed
compact orbits and with the two-dimensional discs attached to the
singular orbits; in the generic case there is just one critical
point on every singular orbit. } \label{reconst}
\end{figure}

 These  pieces of Fermi surface , however, has the holes
with boundaries. They are not the "closed manifolds" anymore. To
get the closed manifolds  let us make the next step:

 We fill in the holes by
the topological $2D$ discs  in the planes orthogonal to ${\bf B}$;
finally we are coming to  the closed surfaces

$${\bar S}_{j} \,\,\, = \,\,\, S_{j} \cup (2D discs)$$
(see Fig. \ref{reconst}).

 This procedure gives  the periodic surface
${\bar S}_{F}$ after the reconstruction and we can define the
"compactified carriers of open trajectories" both in ${\mathbb
R}^{3}$ and ${\mathbb T}^{3}$. Thus we  have  two representations
of the reconstructed Fermi surface:

1) The compact surface without boundary embedded in the space of
quasimomenta ${\mathbb T}^{3}$ (consisting of several pieces
without boundaries);

2) The set of periodic two-dimensional surfaces without boundaries
in the covering space ${\mathbb R}^{3}$.

 Let us formulate now our main intermediate result which was established
using the  theorems extracted from the purely topological
investigations (see, for example \cite{zorich1}).

\vspace{0.5cm}

{\it Let us fix the generic direction of ${\bf B}$ and consider
the set $\{{\bar S}_{j}\}$
carrying the open electron trajectories. Then the only two
situations can be topologically stable with respect to the
small rotations of ${\bf B}$:

(A) The set $\{{\bar S}_{j}\}$ is empty;

(B) The set $\{{\bar S}_{j}\}$ in the torus ${\mathbb T}^{3}$
consists of the even number of surfaces homeomorphic to the
two-dimensional tori ${\mathbb T}^{2}_{j}$; all of them  have the
same homology class in $H_{2}({\mathbb T}^{3})$ up to the sign
(sum of these classes is equal to zero). This property was called
the "Topological resonance". The corresponding representation of
the set $\{{\bar S}_{j}\}$ in total {\bf p}-space ${\mathbb
R}^{3}$ can be described as follows:

 The manifolds ${\bar S}_{j}$ represent the periodically deformed
two-dimensional planes $\Gamma_{(j)\alpha}$ embedded in ${\mathbb
R}^{3}$ with the same common integer-valued mean directions. In
other words we have the set of periodically deformed (warped)
integral planes in ${\mathbb R}^{3}$ which are all parallel in
average and do not intersect each other. This picture remains
unchanged after the small rotation of the magnetic field. }

\vspace{0.5cm}

 The first situation A corresponds to the absence
of the open electron trajectories on the Fermi level. Let us
remind also that we call the  two-dimensional plane "integral" in
${\mathbb R}^{3}$ if it is generated by two reciprocal lattice
vectors.

 The  Topological resonance was first pointed out in
\cite{novmal1,novmal2}. It   plays the crucial role in  the
GSMF-limit as we will see below.

 The topological stability means in particular that the corresponding
picture remains the same after any rotation of ${\bf B}$ small
enough:  the number of connected components as well as the
homological classes of corresponding tori ${\mathbb T}^{2}_{j}$
are the same for all the directions of ${\bf B}$ close enough to
the initial one. Let us make now the important physical conclusion
from our main statement and consider the corresponding corollaries
for the electrical conductivity.

It was also proved  (\cite{dynn7}) that the total measure
of the directions of ${\bf B}$ where different situations can
arise is zero on the unit sphere for the generic Fermi surface
$S_{F}$.

 We will consider these two situations described above as the main
basic  foundation for the total classification of different
regimes in the GSMF-limit for the generic case.

 Let us define now the "Topological numbers" observable in the
Situation B when we really have regular open trajectories.

\vspace{0.5cm}

{\it We call the "Topological numbers" corresponding to the stable
open electron trajectories the triple of integers
$(m_{1},m_{2},m_{3})$ representing the integral 2-plane in the
3-space with reciprocal lattice. (Topologically it is a common
homology class of the 2-tori ${\mathbb T}^{2}_{j}$ in ${\mathbb
T}^{3}$.)}

\vspace{0.5cm}

 This   integers $(m_{1},m_{2},m_{3})$ can be extracted
from common directions of periodically deformed two-dimensional
planes representing $\{{\bar S}_{j}\}$ in ${\mathbb R}^{3}$
with respect to reciprocal lattice $\Gamma^{*}$.
Namely, the planes $\Gamma_{\alpha}$ can be defined
from the equation
$$m^{1}_{\alpha} [{\bf x}]_{1} + m^{2}_{\alpha} [{\bf x}]_{2} +
m^{3}_{\alpha} [{\bf x}]_{3} = 0$$
where $[{\bf x}]_{i}$ are the coordinates in the basis
$\{{\bf g}_{1}, {\bf g}_{2}, {\bf g}_{3}\}$ of the reciprocal
lattice, or equivalently

$$m^{1}_{\alpha} ({\bf x}, {\bf l}_{1}) +
m^{2}_{\alpha} ({\bf x}, {\bf l}_{2}) +
m^{3}_{\alpha} ({\bf x}, {\bf l}_{3}) = 0$$
where $\{{\bf l}_{1}, {\bf l}_{2}, {\bf l}_{3}\}$ is the basis
of the initial lattice in the coordinate space.

 We can formulate now the main statement about the stable open
trajectories in our approach:

\vspace{0.5cm}

{\it All  stable open electron trajectories have the topologically
regular form, i.e. lie in the straight strips of the finite width
in the planes orthogonal to ${\bf B}$ in the ${\bf p}$-space and
pass through them. All  trajectories of this kind have the same
mean directions for the given direction of ${\bf B}$:  in average
they are parallel to each other. The common direction of all these
trajectories is given by the intersection of plane $\Pi({\bf B})$
orthogonal to ${\bf B}$ with some  integral plane
$\Gamma_{\alpha}$ which is locally stable with respect to the
small rotations of ${\bf B}$.}

\vspace{0.5cm}

 The fact that all topologically regular
trajectories are parallel to each other expresses here the
"Topological Resonance" property. It first appeared in
\cite{novmal1,novmal2}. It seems that nothing like that was known
in the classical literature.  In the work \cite{lifpes2} for
example the open electron trajectories with different mean
directions were mistakably demonstrated for some analytic
dispersion relations in the whole regions of the unit sphere
parameterizing directions of ${\bf B}$. We claim however, that
this situation is completely impossible for any open region on the
sphere.
 The important property of topologically regular open trajectories
lies in the following fact:  their contribution to the
conductivity does not differ in the main order in the GSMF-limit
from the (anisotropic) contribution of open periodic trajectories
obtained in the old work \cite{lifazkag}. It is very easy to prove
this statement taking into account that the motion of electron is
linear plus something bounded: one should simply  repeat the
essential arguments of this old work. The "Topological Resonance"
claims more:
  all  trajectories of this kind  give the same
kind of anisotropy in the same coordinate system.  Only this
result makes this behavior experimentally observable. Let us
present here corresponding expressions for the conductivity in the
GSMF-limit for two situations described above.

\vspace{0.5cm}

 {\bf Case I} (Compact trajectories only):

\begin{equation}
\label{sigcltr}
\sigma^{ik} \simeq {n e^{2} \tau \over m^{*}} \,
\left( \begin{array}{ccc}
(\omega_{B}\tau)^{-2} & (\omega_{B}\tau)^{-1} &
(\omega_{B}\tau)^{-1} \cr
(\omega_{B}\tau)^{-1} & (\omega_{B}\tau)^{-2} &
(\omega_{B}\tau)^{-1} \cr
(\omega_{B}\tau)^{-1} & (\omega_{B}\tau)^{-1} & *
\end{array} \right) \,\,\, , \,\,\, \omega_{B}\tau
\rightarrow \infty
\end{equation}

\vspace{0.5cm}

 {\bf Case II} (Open topologically regular trajectories):

\begin{equation}
\label{sigoptr}
\sigma^{ik} \simeq {n e^{2} \tau \over m^{*}} \,
\left( \begin{array}{ccc}
(\omega_{B}\tau)^{-2} & (\omega_{B}\tau)^{-1} &
(\omega_{B}\tau)^{-1} \cr
(\omega_{B}\tau)^{-1} & * & * \cr
(\omega_{B}\tau)^{-1} & * & *
\end{array} \right) \,\,\, , \,\,\, \omega_{B}\tau
\rightarrow \infty
\end{equation}

 Here $\simeq$ means "of the same order in $\omega_{B}\tau$
and $*$ are some constants $\sim 1$. We assume here that the
$z$-axis is always directed along the
magnetic field ${\bf B}$ and the $x$-axis in the plane
$\Pi({\bf B})$ (orthogonal to ${\bf B}$) is directed along the
common mean direction of the topologically regular trajectories
in ${\bf p}$-space in the second case. Let us mention also that
the relations (\ref{sigcltr})-(\ref{sigoptr}) give only the
order of magnitude of $\sigma^{ik}$.

 The anisotropy of the tensor $\sigma^{ik}$ in the formula
(\ref{sigoptr}) gives the experimental possibility of measuring the
mean directions of the topologically regular open orbits for rather
big values of $B$. Using the rotations of the direction of ${\bf
B}$ it is possible also to find the "Stability zone" on the unit
sphere and to determine the corresponding "Topological numbers"
characterizing this stable situation. We see that there is only
one direction ${\hat \eta}$ in the second case where the
conductivity vanishes in the limit $B \rightarrow \infty$.
According to the formula (\ref{sigoptr}) this direction coincides
 with the common mean direction of the topologically regular
trajectories in the ${\bf p}$-space (i.e. orthogonal to the mean
direction of these trajectories in the coordinate ${\bf
x}$-space). The direction ${\hat \eta}({\bf B})$ depends on the
direction of magnetic field. However, it varies in   some integral
plane $\Gamma_{\alpha}$ which is the same for the given "Stability
zone". We can claim that the direction of conductivity decreasing
${\hat \eta} = (\eta_{1}, \eta_{2}, \eta_{3})$ satisfies to the
relation

$$m^{1}_{\alpha} ({\hat \eta}, {\bf l}_{1}) + m^{2}_{\alpha}
({\hat \eta}, {\bf l}_{2}) + m^{3}_{\alpha} ({\hat \eta}, {\bf
l}_{3}) = 0$$ for all the points of stability zone
$\Omega_{\alpha}$ which makes possible the experimental
observation of the numbers $(m^{1}_{\alpha}, m^{2}_{\alpha},
m^{3}_{\alpha})$.

\section{The chaotic cases and their contribution to the
  GSMF-limit.}

 Let us say now some words about the chaotic trajectories which can
arise in the special cases for rather complicated Fermi surfaces.
You have to remember that they can appear only for the zero
measure set of directions of the magnetic field. We think that for
the generic Fermi surfaces the fractal (or Hausdorf) dimension of
this set is strictly less than 1 (it was certainly proved by
Dynnikov that it is no more than 1 for the generic Fermi surfaces,
but it can be more for the nongeneric ones--see numerical studies
in the works \cite{dynn7,DeLeo}. Anyway, there is no proof of this
until now.

 We will first mention Tsarev's example of weakly chaotic trajectory
having an asymptotic direction in ${\mathbb R}^{3}$
(\cite{tsarev}). We will not describe here the details of
corresponding Fermi surface (see \cite{malnov4}) and just say the
trajectory of this kind can not be imbedded in any straight strip
of finite width in ${\bf p}$-space. However this trajectory has
always asymptotic direction in the plane orthogonal to ${\bf B}$.
The motion is linear plus smaller (but unbounded) terms. We can
always choose the coordinate system such that the average values
of the group velocities satisfy to the following condition:

$$\langle v^{x}_{gr} \rangle = 0 \,\,\, , \,\,\,
\langle v^{y}_{gr} \rangle \neq 0 \,\,\, , \,\,\,
\langle v^{z}_{gr} \rangle \neq 0 $$

 Here again the $z$-axis coincides with the direction of
${\bf B}$ and the $x$-axis is directed along the asymptotic
direction of the chaotic trajectory in ${\bf p}$-space.

 The corresponding behavior of conductivity in GSMF-limit
does not coincide completely with the formula (\ref{sigoptr}),
however the following formulae for the behavior of
$\sigma^{ik}(B)$ can be proved:

\begin{equation}
\label{sigtsar}
\sigma^{ik}({\bf B}) \simeq {n e^{2} \tau \over m^{*}} \,
\left( \begin{array}{ccc}
o(1) & o(1) & o(1) \cr
o(1) & * & * \cr
o(1) & * & *
\end{array} \right) \,\,\, , \,\,\, \omega_{B} \tau \rightarrow
\infty
\end{equation}
which replaces the formula (\ref{sigoptr}) for the case of weakly
chaotic   trajectories. Let us omit here all details of the weakly
 chaotic trajectories and just point out that the asymptotic
direction of the weakly chaotic trajectory  can be also observed
experimentally due to the same reasons as in the case of
topologically regular trajectories. However, unlike the
topologically regular case, the weakly chaotic trajectories  are
unstable with respect to generic small rotations of ${\bf B}$.
They  correspond to some very small sets on the unit sphere. At
last we say that the trajectories of this kind can appear only for
the direction of ${\bf B}$ of  irrationality 2, i.e. the plane
$\Pi({\bf B})$ should contain one reciprocal lattice vector in
this situation.

 Let us say now some words about more general strongly chaotic
trajectories  which do not have any asymptotic direction in
${\mathbb R}^{3}$. We will not describe here the corresponding
construction (see \cite{dynn4}) and just give the main features of
such trajectories.

 First of all,
these trajectories can arise only in the case of magnetic field of
irrationality 3 and the corresponding carriers have then the genus
$\geq 3$. This kind of trajectories are completely unstable with
respect to the small rotations of ${\bf B}$ and can be observed
just for special fixed directions of ${\bf B}$ in the case of
rather complicated Fermi surfaces. The approximate form of some
trajectories of this kind is shown at Fig. \ref{regandch}, b.
Moreover, if the genus of  Fermi surface is not very high ($< 6$)
it can always be stated that the corresponding carrier of open
strongly chaotic trajectories is invariant under the involution
${\bf p} \rightarrow - {\bf p}$ (after the appropriate choice of
the initial point in ${\mathbb T}^{3}$). The ergodic theorem
applied to the open trajectories on the carrier gives then
immediately the relations:

$$\langle v^{x}_{gr} \rangle = 0 \,\,\, , \,\,\, \langle
v^{y}_{gr} \rangle = 0 \,\,\, , \,\,\, \langle v^{z}_{gr} \rangle
= 0 $$ for all three components of the group velocity on any of
such trajectories. This important fact leads to the rather
non-trivial behavior of corresponding contribution to the
conductivity for $B \rightarrow \infty$. Namely we can show that 
all the components of the corresponding contribution to 
$\sigma^{ik}({\bf B})$ become actually zero in the limit 
$B \rightarrow \infty$
(\cite{maltsev}). We can write then for this contribution:

\begin{equation}
\label{sigdyn}
\sigma^{ik}({\bf B}) \simeq {n e^{2} \tau \over m^{*}} \,
\left( \begin{array}{ccc}
o(1) & o(1) & o(1) \cr
o(1) & o(1) & o(1) \cr
o(1) & o(1) & o(1)
\end{array} \right)
\end{equation}
for $B \rightarrow \infty$.\footnote{Actually the component
$\sigma^{zz}({\bf B})$ contains the non-vanishing term of order
of $T^{2}/\epsilon_{F}^{2}$ for $B \rightarrow \infty$ for
non-zero temperatures (\cite{maltsev}). However, this parameter
is very small for the normal metals and we don't take it here
in the account.}

 We see that the strongly chaotic trajectories (''Dynnikov type'')
do not give any contribution even for conductivity along the
magnetic field ${\bf B}$ for rather big values of $B$. In the work
\cite{maltsev} also the special "scaling" asymptotic behavior of
$\sigma^{ik}({\bf B})$ were suggested. Let us note, however, that
the full conductivity tensor include also the contribution of
compact (closed) electron trajectories having the form
(\ref{sigcltr}) which presents in general as the additional
contribution in all the cases described above. We can so claim
that the strongly chaotic behavior does not  remove completely
the conductivity along the magnetic field ${\bf B}$ because of the
contribution of compact trajectories. However, the sharp local
minimum in this conductivity can still be observed in this case
since a part of the Fermi surface will be effectively excluded
from the conductivity in this situation.

 It can be proved (see \cite{dynn7}) that for the generic Fermi
surfaces the measure of directions of magnetic field ${\bf B}$
where the strongly chaotic behavior  can be found on the Fermi
surface is equal to zero. However, the total set on the unit
sphere corresponding to the strongly chaotic trajectories of this
kind can be rather complicated set with the non-trivial Hausdorf
dimension. We expect that the Hausdorf dimension of this set is
strictly less than 1 for the generic Fermi surfaces. For the
nongeneric cases it might be even more than 1.

 At last let us say that we expect that either  the small stability zones or
 the  strongly  chaotic trajectories in fact were observed
 in the experimental data
represented in \cite{gaid} (see \cite{maltsev,novmal2}). However
these data are not detailed enough  (for example the conductivity
along magnetic field was not measured in this experiments).

\vspace{0.5cm}

 Let us describe now the total picture for the angle diagram
of conductivity in normal metal in the case of geometric
strong magnetic field limit (\cite{malnov3,malnov4}.
Namely, we can observe the following objects on the
unit sphere parameterizing the directions of ${\bf B}$:

 1) The "stability zones" $\Omega_{\alpha}$ corresponding to
topologically regular open trajectories and parameterized by
some integral planes $\Gamma_{\alpha}$ in the reciprocal lattice
("Topological Numbers"). All the "stability zones" have the
piecewise smooth boundaries on $S^{2}$.

 The corresponding behavior of conductivity
is described by the formula (\ref{sigoptr}) and reveals the strong
anisotropy in the planes orthogonal to the magnetic field. For
rather complicated Fermi surfaces we can observe also the
"sub-boundaries" of the stability zones where the coefficients in
(\ref{sigoptr}) can have the sharp "jump" but do not change the
"Topological Numbers" characterizing the "Stability zone"
$\Omega_{\alpha}$.

\vspace{0.5cm}

 2) The net of the one-dimensional curves containing directions
of irrationality $\leq 2$ where the additional periodic open
trajectories in ${\bf p}$-space can appear. The corresponding
parts of the net are always the parts of the big (passing
through the center of $S^{2}$) circles orthogonal to some
reciprocal lattice vector. The asymptotic behavior of conductivity is
given again by the formula (\ref{sigoptr}) but unstable with
respect to the small rotations of ${\bf B}$ going out from the
corresponding curves.

\vspace{0.5cm}

 3) The "Special rational directions".

 We call the special rational direction the direction of
${\bf B}$ orthogonal to the integral plane $\Gamma_{\alpha}$
corresponding to some stability zone $\Omega_{\alpha}$
in case when this direction belongs to the same stability
zone on the unit sphere. We don't discuss here all the
specific features which can appear in this situation and just
say that some specialties can arise here. Let us give here
the reference on the papers \cite{malnov3,malnov4} where
all corresponding possibilities are discussed.

\vspace{0.5cm}

 4) The weakly chaotic open orbits
(${\bf B}$ of irrationality $2$).

 We can have points on the unit sphere where the open orbits
are weakly chaotic. All open trajectories still have the
asymptotic direction in this case and the conductivity reveals the
strong anisotropy in the plane orthogonal to ${\bf B}$ as $B
\rightarrow \infty$. The $B$ dependence, however is slightly
different from the formula (\ref{sigoptr}) in this case.

\vspace{0.5cm}

 5) The strongly chaotic open orbits
(${\bf B}$ of irrationality $3$).

 For some points on $S^{2}$ we can have the strongly chaotic open
orbits on the Fermi surface. At these points the local minimum of
conductivity along the magnetic field is expected. The
conductivity along ${\bf B}$ however remains finite as $B
\rightarrow \infty$ in general situation because of the
contribution of compact trajectories.

\vspace{0.5cm}

 6) At last we can have the open regions on the unit sphere
where only the compact trajectories on the Fermi level are
present (Situation A). The asymptotic behavior of conductivity
tensor is given then by the formula (\ref{sigcltr}).

\vspace{0.5cm}

\begin{figure}
\begin{center}
\epsfig{file=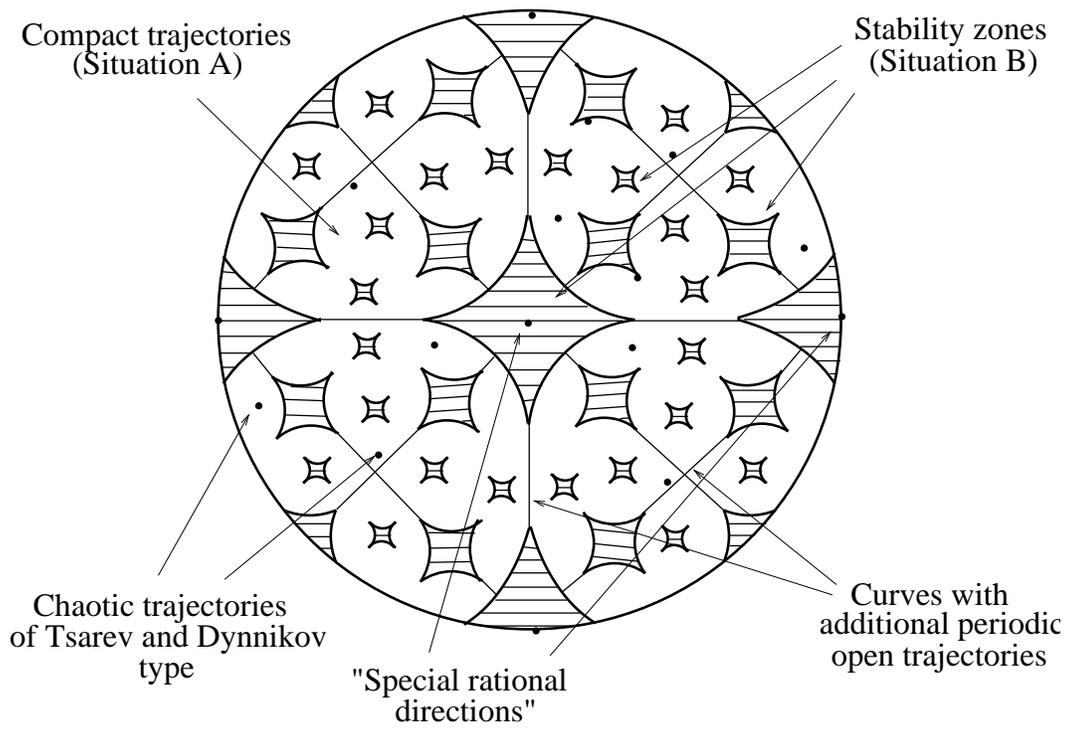,width=14.0cm,height=9.5cm}
\end{center}
\caption{The schematic representation of possible regimes
for the different directions of the magnetic field ${\bf B}$
on the unit sphere.
}
\label{classif}
\end{figure}

 At the Fig. \ref{classif} we show the schematic picture of
the regimes described above for different directions of magnetic
field ${\bf B}$.

 Let now point out some new features connected with the
"magnetic breakdown" (self-intersecting Fermi surfaces) which
can be observed for rather strong magnetic fields. Up to this
point it has been assumed throughout that different parts
of the Fermi surface do not intersect with each other.
However, it is possible for some special lattices that the
different components of the Fermi surface (parts corresponding
to different conductivity bands) come very close to each other
and may have an effective "reconstruction" as a result of
the "magnetic breakdown" in strong magnetic field limit.
In this case we can have the situation of the electron
motion on the self-intersecting Fermi surface such that   
the intersections with other pieces do not affect at all the
motion on one component. (The physical conditions for the
corresponding values of $B$ can be found in \cite{etm}).
In this case the picture described above should be considered
independently for all the non-selfintersecting pieces
of Fermi surface and we can have simultaneously several      
independent angle diagrams of this form on the unit sphere.


\begin{thebibliography}{99}

\bibitem{lifazkag} I.M.Lifshitz, M.Ya.Azbel, M.I.Kaganov.
{\it Sov. Phys. JETP} {\bf 4}, 41 (1957).

\bibitem{lifpes1} I.M.Lifshitz, V.G.Peschansky.
{\it Sov. Phys. JETP} {\bf 8}, 875 (1959).

\bibitem{lifpes2} I.M.Lifshitz, V.G.Peschansky.
{\it Sov. Phys. JETP} {\bf 11}, 137 (1960).

\bibitem{gaid} Yu.P.Gaidukov.
{\it Sov. Phys. JETP} {\bf 10}, 913 (1960).

\bibitem{lifkag1} I.M.Lifshitz, M.I.Kaganov.
{\it Sov. Phys. Usp.} {\bf 2}, 831 (1960).

\bibitem{lifkag2} I.M.Lifshitz, M.I.Kaganov.
{\it Sov. Phys. Usp.} {\bf 5}, 411 (1962).

\bibitem{kittel} C. Kittel. Quantum Theory of Solids.
John Wiley \& Sons, Inc. NewYork - London, 1963. 

\bibitem{etm} I.M.Lifshitz, M.Ya.Azbel, M.I.Kaganov.
Electron Theory of Metals. Moscow, Nauka, 1971.
Translated: New York: Consultants Bureau, 1973.

\bibitem{ziman} J.M. Ziman. Principles of the Theory of Solids.
Cambridge, At the University Press, 1972.

\bibitem{abrikosov} A.A.Abrikosov.
Fundamentals of the Theory of Metals.
"Nauka", Moscow (1987). Translated:
Amsterdam: North-Holland, 1998.

\bibitem{novikov1} S.P.Novikov.
{\it Russian Math. Surveys} {\bf 37}, 1 (1982).

\bibitem{zorich1} A.V.Zorich.
{\it Russian Math. Surveys} {\bf 39}, 287 (1984).

\bibitem{novikov2} S.P.Novikov.
Proc. Steklov Inst. Math. 1 (1986).

\bibitem{arnold} V.I.Arnold. {\it Functional analysis and
its appilcations} {bf 25}:2 (1991).

\bibitem{sinkhan} Ya.G.Sinai, K.M.Khanin.
{\it Functional analysis and its appilcations}
{bf 26}:3 (1992).

\bibitem{dynn1} I.A.Dynnikov.
{\it Russian Math. Surveys} {\bf 57}, 172 (1992).

\bibitem{dynn2} I.A.Dynnikov.
{\it Russian Math. Surveys} {\bf 58} (1993).

\bibitem{dynn3} I.A.Dynnikov. "A proof of Novikov's conjecture
on semiclassical motion of electron."
{\it Math. Notes} {\bf 53}:5, 495 (1993).

\bibitem{tsarev} S.P.Tsarev.
Private communication. (1992-93).

\bibitem{novikov3} S.P.Novikov. "Quasiperiodic structures in topology".
Proc. Conference "Topological Methods in Mathematics",
dedicated to the 60th birthday of J.Milnor, June 15-22, S.U.N.Y.
Stony Brook, 1991. Publish of Perish, Houston, TX, pp. 223-233 (1993).

\bibitem{zorich2} A.V.Zorich.
Proc. "Geometric Study of Foliations" (Tokyo, November 1993)/
ed. T.Mizutani et al. Singapore: World Scientific, 479-498
(1994).

\bibitem{novikov4} S.P.Novikov. Proc. Conf. of Geometry,
December 15-26, 1993, Tel Aviv University (1995).

\bibitem{novmal1} S.P.Novikov, A.Ya.Maltsev.
{\it ZhETP Lett.} {\bf 63}, 855 (1996).

\bibitem{dynn5} I.A.Dynnikov.
"Surfaces in 3-Torus: Geometry of plane sections."
Proc.of ECM2, BuDA, 1996.

\bibitem{dynn4} I.A.Dynnikov.
"Semiclassical motion of the electron. A proof of the Novikov
conjecture in general position and counterexamples."
American Mathematical Society Translations, Series 2, Vol. 179,
Advances in the Mathematical Sciences. Solitons, Geometry, and
Topology: On the Crossroad. Editors: V.M.Buchstaber,
S.P.Novikov. (1997)

\bibitem{dynmal} I.A.Dynnikov, A.Ya.Maltsev.
{\it ZhETP} {\bf 85}, 205 (1997).

\bibitem{maltsev} A.Ya. Maltsev.
{\it ZhETP} {\bf 85}, 934 (1997).

\bibitem{novmal2} S.P.Novikov, A.Ya.Maltsev.
{\it Physics-Uspekhi} {\bf 41}(3), 231 (1998).

\bibitem{dynn7} I.A.Dynnikov.
{\it Russian Math. Surveys} {\bf 54}, 21 (1999).

\bibitem{novikov5} S.P.Novikov.
{\it Russian Math. Surveys} {\bf 54}:3, 1031 (1999).

\bibitem{DeLeo} R.D.Leo. SIAM Journal on Applied Dynamical Systems,
{\bf 2}:4, 517-545 (2003).

\bibitem{malnov3} A.Ya.Maltsev, S.P.Novikov,
ArXiv: math-ph/0301033,
Bulletin of Braz. Math. Society, New Series {\bf 34} (1), 171-210 
(2003).

\bibitem{malnov4} A.Ya.Maltsev, S.P.Novikov.//
ArXiv: cond-mat/0304471

\bibitem{maltsev2}  A.Ya. Maltsev, Arxiv: cond-mat/0302014

\bibitem{PanSpTeuf1} G. Panati, H. Spohn, S. Teufel. Phys. Rev. Lett.
{\bf 88}, 250405 (2002).

\bibitem{PanSpTeuf2} G. Panati, H. Spohn, S. Teufel. //arXiv:
math-ph/0212041







\end{thebibliography}
\end{document}